# Talk&Learn: Improving Conversation Experience and Creating Opportunities for Foreign Language Learning

Yaohua Xie

Institute of Software, Chinese Academy of Sciences

Yaohua.Xie@hotmail.com; fjpnxyh2000@163.com

**INTRODUCTION**

Nowadays, language barrier is still a big challenge despite of the greater and greater demand of cross-cultural communication[1]. On the one hand, people wish to communicate with others in different languages with the help of certain tools or services. On the other hand, many people wish they can talk freely in other languages by themselves eventually. However, existing Real-Time Translation Interfaces (RTTI)[2] do not provide experience as natural and efficient as monolingual communication. Also, such systems do not provide functions supporting language learning. This results in the waste of both time and potential language context.

In order to overcome the above limitations, we propose a solution named "**Talk&Learn**". Its **core idea** is to rearrange ("**Delay-Match**") the real-time videos and translated texts or speeches, so as to gain better naturalness and efficiency. At the same time, this will **create extra free time** for users. Inspired by "Wait-Learning"[3] etc., we propose to utilize the free time for contextual **language learning**.

**LITERATURE REVIEW**

After decades of research, automatic Spoken Language Translation (SLT) is becoming more and more mature. Some researchers believe that it is usable in certain fields[4]. RTTI is the interface between language translation service and the users. Existing RTTIs usually show real-time video streams to users, then wait for translation, and finally display translated texts or play translated speeches[5, 6]. In most solutions, visual components such as facial expressions, eye gazes and gestures, are utilized to make communication easier. The combination of these components and translation technique is beneficial to user satisfaction and sense of spontaneity in conversation[2]. Text and speech have also been proven to be helpful for non-native speakers[7]. Researches in the field of second language learning have demonstrated that closed captions can improve participants' comprehension of foreign language when seeing DVD videos[8]. Such "batch mode captions" are shown line by line, and may contain the speaker's future words before being spoken in the video[9]. Therefore, people did not think such situation can happen in real-time conversations[9]. Similarly, translated speeches also appear after corresponding videos.

In order to solve these problems, we propose the "Delay-Match" approach to synchronize videos and texts/speeches. Meanwhile, this approach also create extra "free time", which we further propose to use for language learning. In addition, the problem of privacy could also be diminished by such a solution. In order to protect privacy, FocalSpace uses synthetic blur effects to diminish the background [10]. In contrast, our solution hides "irrelevant" videos, e.g., when users are waiting expressionlessly for translation results. These periods of time are utilized to learn previous speeches or texts.

Many researches have demonstrated the value of contextual micro-learning[11, 12]. Various media have been employed such as instant messages[3], web pages[11], Facebook feeds[13] and live wallpapers[14]. For example, in ALOE prototype, selected sets of English words are dynamically replaced with their foreign translations, so that users can learn vocabularies during web browsing[11]. We propose to adopt the integration of video, audio and text as the medium, and help users learn foreign words, phrases or sentences during listening, speaking and reading. Showing the history of translation not only could help users to comprehend long messages[1], but also provide the chance to review, practice and test their foreign language just-in-time while communicating. As 3D space provide more room for information visualization, we will try using 3D timeline to organize conversation history.

**METHODS**

In **existing applications** such as Skype Translator and Vocre, the user **cannot understand** what he/she hears when receiving the original video from the other one. Then he/she **must wait** for the translated result, ether in text or speech. Finally, he/she reads the text or listens to the speech, **without the help** of facial expression. The user's **brain is busy** in most time of the above process, but the information he/she gets is **not synchronous**. We **propose** to **delay** the presentation of videos and **match** them with the translated texts or speeches. The translated texts can be shown as "**batch mode captions**" which used to be not available in video meeting[9]. The translated speeches can also replace the original ones in the video. The output of this process is called "**synthesized videos**" here. By doing so, the user can get **natural perception** of the other's talk. Furthermore, the above "see" and "wait" stages become **free time** for the user. That creates a suitable room for **online education**, e.g., **contextual micro-learning**.

We **propose** to **learn** foreign language during the **free time** using the **conversation history**. Typically, the user can learn from the translated texts or speeches of the previous messages he/she has **got**, or the translated texts of the previous messages he/she has **sent**. Also, the user can choose to send messages in foreign language (**need no translation**), to see whether the other user can understand. Or, he/she can choose to speak a sentence **repeating or imitating** one of the previous ones, and **not send it out (just for practice)**. During the use, the user may be able to speak **more and more** sentences without translation. If the user can get extra benefit from these efforts, e.g., the **discount** of translation fees, he/she could be more **motivated**. Besides, the **percentage** of the messages **without translation** can be calculated and analyzed, so as to provide a feedback and encouragement to the user. The **percentage** of the messages translated **by machine** may also be analyzed, so as to reflect the ability of automatic translation. After a period of time, the users may gradually be able to communicate by **themselves**, and the translation system become **more automatic** while the corpus becomes larger. As a result, the **cost** of both the user and translation service provider could become **less and less**, and the **technique** of machine translation could become more and more **mature**.

In **existing solutions**, users' videos are **always displayed** to each other. But the users are not showing meaningful facial expressions, and even seem **awkward** during waiting periods. We **propose** to **only send the synthesized videos**, which match the user's facial expression and the translated text or speech. **Auxiliary pictures** are shown as hints to the user in the rest stages, e.g., when the other one is speaking or when the system is translating. This may provide more **freedom and privacy** for the users to do multiple tasks, and also make the conversation **less awkward**. Because the videos are delayed, the judgement of the above stages can be **performed automatically** by video/speech analysis. Through certain **indicator**, the user is able to know whether he/she is "visible" to the other one. For example, a small window appears when the user is "visible", showing his/her own picture. Besides, he/she can intentionally choose to be "visible" at any time when necessary.

All messages are displayed as pictures along a **timeline** in a virtual **3D space**. The user can **drag** the pictures to **zoom in** any of them, so as to **review** conversation and learn language.

The purpose of **Talk&Learn** is to firstly perform actual conversations, and secondly learn languages. Therefore, the following two aspects should be evaluated:

Brief proposal

**Efficacy of conversation**: Can users communicate effectively using the system?

**Efficacy of learning**: Can users improve foreign language through conversations?

In order to evaluate, the participants need to perform a story telling task[1]. They should generate a story about a given topic using free form speech similar to a normal conversation. There are some keywords serve as hints, which may make it easier for the participants to keep a conversation going.

During the waiting periods, the participants should review, repeat, retell or recite previous messages. They are encouraged to make sentences imitating those they have already learned during conversation. They are also encouraged to use foreign language as possible as they can. Extra reward may be set for those who correctly use foreign language more often.

After several days of use, the participants need to take a language test. They should try to retell (in foreign language) or recognize the sentences appeared in their conversations. The aim is to check how many sentences they have learned during the conversations.

## RESULTS

The following data should be collected in order to conduct post-hoc analysis:

**Conversation log**: Including a timeline and all the events attached to it. The events indicate what have happened during the conversation, i.e., (1) videos and audios of original speeches, (2) transcribed texts, (3) translated texts and (4) translated speeches and (5) synthesized videos. The start time and end time of each event should be recorded.

**Questionnaire**: In order to evaluate the system subjectively, the users should fill in questionnaires on communication efficacy and learning efficacy. Each questionnaire is comprised of Likert scale questions and an optional free-text response. The users can judge how well they have understood each other during the conversations, the perceived improvement of their foreign language, etc. They can also comment on their opinion and suggestions on the usage, interface and design of the system.

**Test results**: After conversations, the users need to take a test so as to check the improvement of their language. Their answers are recorded when retelling or recognizing the messages.

**Video recording**: During each part of the test, the whole scene should be recorded with a camcorder. This will provide materials for a post-hoc video observation.

Based on the above data, analysis could be conducted on the following aspects:

- Naturalness of conversation: How natural the participant think their conversations are.
- Effectiveness of conversation: How effective the participant think their conversations are.
- Difficulty of understanding: How difficult the participant think they understand each other.
- Difficulty of turn taking: How difficult the participant think turn taking is.
- Number of retold sentence: How many sentences the participant can retell.
- Number of recognized sentence: How many sentences the participant can recognize.
- Learning efficacy: How good the participant think the efficacy of learning is.
- Disruption: How severe the participant think conversations are disrupted by learning.
- Preference of usage manner: How the participant like this manner of language learning.

## DISCUSSION

Here we introduce the idea of "Delay-Match" to create "Synthesized Videos" for real-time multilingual communication, and utilize the resulting free time for contextual language learning. This may not only make conversation more effective, but also make better use of the time and context. It is also beneficial to translation service providers and the development of corpus. This technique is suitable for both desktop applications and mobile ones. It is also independent of the specific implementation of translation system. In the future, more research should be done to better integrate learning process into conversation.